\documentclass[twoside,12pt]{article}
\usepackage{epsfig}

\def\Journal#1#2#3#4{{#1} {#2} (#4) #3 }

\def\NPA{{\em Nucl. Phys.} A}

\def\PRD{{\em Phys. Rev.} D}
\def\PRC{{\em Phys. Rev.} C}

\def\ZPA{{\em Z. Phys.} A}

\def\r{\vec r}

\newcommand{\be}{\begin{equation}}
\newcommand{\ee}{\end{equation}}
\newcommand{\bea}{\begin{eqnarray}}
\newcommand{\eea}{\end{eqnarray}}

 \def\bean{\begin{eqnarray*}}
 \def\eean{\end{eqnarray*}}

 \def\l{\left}
 \def\r{\right}

 \def\bm#1{\mbox{\boldmath$#1$}}
 
 \def\ksim{\mathrel{\rlap{\lower0.2em\hbox{$\sim$}}\raise0.2em\hbox{$<$}}}

\begin{document}
\title{\vspace{1cm} Parton dynamics and hadronization from the sQGP}

\author{W.\ Cassing$^1$,
E. L.\ Bratkovskaya$^{2}$, Y.-.Z. Xing$^{3}$
\\ \\ $^1$Institut f\"{u}r Theoretische Physik,
   Universit\"{a}t Giessen,  Germany\\
$^2$FIAS and
      JWG Universit\"{a}t Frankfurt,  Frankfurt am Main, Germany\\
      $^3$Dept. of Physics, Tianshui Normal University, P. R. China}
   \date{}
\maketitle

\begin{abstract}
  The hadronization of an expanding partonic fireball  is
studied within the Parton-Hadron-Strings Dynamics (PHSD) approach
which is based on a dynamical quasiparticle model (DQPM) matched
to reproduce lattice QCD results in thermodynamic equilibrium.
Apart from strong parton interactions the expansion and
development of collective flow is found to be driven by strong
gradients in the parton mean-fields. An analysis of the elliptic
flow $v_2$ demonstrates a linear correlation with the spatial
eccentricity $\epsilon$ as in case of ideal hydrodynamics. The
hadronization occurs  by quark-antiquark fusion or 3 quark/3
antiquark recombination which is described by  covariant
transition rates. Since the dynamical quarks become very massive,
the formed resonant 'pre-hadronic' color-dipole states ($q\bar{q}$
or $qqq$) are of high invariant mass, too, and sequentially decay
to the groundstate meson and baryon octets increasing the total
entropy. This solves the entropy problem in hadronization in a
natural way. Hadronic particle ratios turn out to be in line
with those from a grandcanonical partition function at temperature
$T \approx 170$ MeV.
 \end{abstract}

%\pacs{25.75.-q, 13.60.Le, 14.40.Lb, 14.65.Dw}
\maketitle

\section{Introduction}

The 'Big Bang' scenario implies that in the first micro-seconds of
the universe the entire state has emerged from a partonic system
of quarks, antiquarks and gluons -- a quark-gluon plasma (QGP) --
to color neutral hadronic matter consisting of interacting
hadronic states (and resonances) in which the partonic degrees of
freedom are confined. The nature of confinement and the dynamics
of this phase transition has motivated a large community for
several decades  and is still an outstanding question of todays
physics. Early concepts of the QGP were guided by the idea of a
weakly interacting system of partons which might be described by
perturbative QCD (pQCD). However, experimental observations at the
Relativistic-Heavy-Ion Collider (RHIC) indicated that the new
medium created in ultrarelativistic Au+Au collisions was
interacting more strongly than hadronic matter  and consequently
this concept had to be given up. Moreover, in line with
theoretical studies in Refs. \cite{Shuryak,Thoma,Andre} the medium
showed phenomena of an almost perfect liquid of partons
\cite{STARS} as extracted from the strong radial expansion
and elliptic flow of hadrons \cite{STARS}.

The question about the  properties of this (nonperturbative) QGP
liquid is discussed controversially in the literature  and
dynamical concepts describing the formation of color neutral
hadrons from partons are scarce \cite{Bleicher,Koal1,Koal2,Biro}.
A fundamental issue for hadronization models is the conservation
of 4-momentum as well as the entropy problem because by
fusion/coalescence of massless (or low constituent mass) partons
to color neutral bound states of low invariant mass (e.g. pions)
the number of degrees of freedom and thus the total entropy is
reduced in the hadronization process \cite{Koal1,Koal2}. This
problem - a violation of the second law of thermodynamics  as well
as of the conservation of four-momentum and flavor currents -
definitely needs a sound dynamical solution.

A consistent dynamical approach - valid also for strongly
interacting systems - can be formulated on the basis of
Kadanoff-Baym (KB) equations \cite{KBaym,Sascha1} or off-shell
transport equations in phase-space representation, respectively
\cite{Sascha1,Juchem}. In the KB theory the field quanta are
described in terms of propagators with complex selfenergies.
Whereas the real part of the selfenergies can be related to
mean-field potentials, the imaginary parts  provide information
about the lifetime and/or reaction rates of time-like 'particles'
\cite{Andre}. Once the proper (complex) selfenergies of the
degrees of freedom are known the time evolution of the system is
fully governed  by off-shell transport equations (as described in
Refs. \cite{Sascha1,Juchem}).

The determination/extraction of complex selfenergies for the
partonic degrees of freedom has been performed in Refs.
\cite{Andre,Cassing06,Cassing07} by fitting lattice QCD (lQCD)
'data' within  the Dynamical QuasiParticle Model (DQPM). In fact,
the DQPM allows for a simple and transparent interpretation of
lattice QCD results for thermodynamic quantities as well as
correlators and leads to effective strongly interacting partonic
quasiparticles with broad spectral functions.

\section{The PHSD approach}

The Parton-Hadron-String-Dynamics (PHSD) approach is a microscopic
covariant transport model that incorporates effective partonic as
well as hadronic degrees of freedom and involves a dynamical
description of the hadronization process from partonic to hadronic
matter \cite{CasBrat}. Whereas the hadronic part is essentially
equivalent to the conventional Hadron-Strings-Dynamics (HSD)
approach \cite{HSD} the partonic dynamics is based on the
Dynamical QuasiParticle Model (DQPM) \cite{Cassing06,Cassing07}
which describes QCD properties in terms of single-particle Green's
functions (in the sense of a two-particle irreducible (2PI)
approach).

We briefly recall the basic assumptions of the DQPM: Following Ref.
\cite{Andre05} the dynamical quasiparticle mass (for gluons and quarks) is
assumed to be given by the thermal mass in the asymptotic
high-momentum regime, which is proportional to the coupling
$g(T/T_c)$ and the temperature  $T$ with a running coupling
(squared),
\begin{eqnarray}
 g^2(T/T_c) = \frac{48\pi^2}{(11N_c - 2 N_f)
\ln[\lambda^2(T/T_c-T_s/T_c)^2}]\ .
 \label{eq:g2}
\end{eqnarray}
Here $N_c = 3$ stands for the number of colors while $N_f$ denotes
the number of flavors. The parameters controlling the infrared
enhancement of the coupling $\lambda = 2.42$ and $T_s = 0.46 T_c$
have been fitted in \cite{Andre05} to lQCD results for the entropy
density $s(T)$. An almost perfect reproduction of the energy
density $\varepsilon(T)$ and the pressure $P(T)$ from lQCD is
achieved as well.

The width  for gluons and quarks (for vanishing chemical potential
$\mu_q$) is adopted in the form
\begin{eqnarray} \label{eq:gamma}
  \gamma_g(T)
  =
  \frac{3 g^2 T}{8 \pi} \,  \ln\left( \frac{2c}{g^2}\right)  \, , \
    \gamma_q(T)
  =
   \frac{g^2 T}{6 \pi} \,  \ln
  \left(\frac{2c}{g^2}\right)
  \, ,
\end{eqnarray}
where $c=14.4$ (from Ref. \cite{Andre}) is related to a magnetic
cut-off. We stress that a non-vanishing width $\gamma$ is the
central difference of the DQPM to conventional quasiparticle
models \cite{qp1}. It influence is essentially seen in correlation
functions as e.g. in the stationary limit of the correlation
function in the off-diagonal elements of the energy-momentum
tensor $T^{kl}$ which defines the shear viscosity $\eta$ of the
medium \cite{Andre}. Here a sizable width is mandatory to obtain a
small ratio in the shear viscosity to entropy density $\eta/s$.

In line with \cite{Andre05}
the parton spectral functions thus are no longer $\delta-$ functions in the
invariant mass squared but taken as
\begin{eqnarray}
 \rho_j(\omega)
 =
 \frac{\gamma_j}{ E_j} \l(
   \frac{1}{(\omega-E_j)^2+\gamma_j^2} - \frac{1}{(\omega+E_j)^2+\gamma_j^2}
 \r)
 \label{eq:rho}
\end{eqnarray} separately for quarks and gluons ($j=q,\bar{q},g$).
With the convention $E^2(\bm p) = \bm p^2+M_j^2-\gamma_j^2$, the
parameters $M_j^2$ and $\gamma_j$ are directly related to the real
and imaginary parts of the  retarded self-energy, e.g. $\Pi_j =
M_j^2-2i\gamma_j\omega$.
With the spectral functions fixed by Eqs.  (1)-(3) the total energy
density in the DQPM (at vanishing quark chemical potential) can be
evaluated as
\begin{equation} \label{ener}
T^{00} = d_g \int_0^\infty  \frac{d\omega}{2 \pi}
\int \frac{d^3 p}{(2 \pi)^3}\ 2 \omega^2 \rho_g(\omega, {\bf p})
n_B(\omega/T)
 + d_q \int_0^\infty  \frac{d\omega}{2 \pi}
\int \frac{d^3 p}{(2 \pi)^3} \ 2 \omega^2 \rho_q(\omega, {\bf p})
n_F(\omega/T) \ , \end{equation}  where $n_B$ and $n_F$ denote the
Bose and Fermi functions, respectively. The number of transverse
gluonic degrees of freedom is $d_g=16$ while the fermic degrees of
freedom amount to $d_q=4 N_c N_f=36$ in case of three flavors
($N_f$=3). The pressure $P$ then may be obtained by integrating
the differential thermodynamic relation \begin{equation} P - T
\frac{\partial P}{\partial T}= - T^{00} \end{equation} with the
entropy density $s$ given by
\begin{equation}
s = \frac{\partial P}{\partial T} = \frac{T^{00}+P}{T} \ .
\end{equation}
This approach is thermodynamically consistent and represents a 2PI
approximation to hot QCD (once the free parameters in (1) and (2) are
fitted to lattice QCD results as in Refs.
\cite{Andre,Cassing06,Cassing07}).

As outlined in detail in Refs. \cite{Cassing06,Cassing07} the energy
density functional (\ref{ener}) can be separated in space-like and
time-like sectors when the spectral functions aquire a finite width.
The space-like part of (\ref{ener}) defines a potential energy density
$V_p$ since the field quanta involved are virtuell and correspond to
partons exchanged in interaction diagrams. The time-like part of
(\ref{ener}) corresponds to effective field quanta which can be
propagated within the light-cone. Related separations can be made for
virtuell and time-like parton densities \cite{Cassing06,Cassing07}.
Without repeating the details we mention that mean-field potentials for
partons can be defined by the derivative of the potential energy
density $V_p$ with respect to the time-like parton densities and
effective interactions by second derivatives of $V_p$ (cf. Section 3 in
Ref. \cite{Cassing07}).

Based on the DQPM we have developed an off-shell transport
approach \cite{CasBrat} denoted as PHSD where the
degrees-of-freedom are dynamical quarks, antiquarks and gluons
($q, \bar{q}, g$) with rather large masses and broad spectral
functions in line with (1) - (3) as well as the conventional
hadrons (described in the standard HSD approach \cite{HSD}). On
the partonic side the following elastic and inelastic interactions
are included $qq \leftrightarrow qq$, $\bar{q} \bar{q}
\leftrightarrow \bar{q}\bar{q}$, $gg \leftrightarrow gg$, $gg
\leftrightarrow g$, $q\bar{q} \leftrightarrow g$  exploiting
'detailed-balance' with interaction rates from the DQPM. The
hadronisation, i.e. transition from partonic to hadronic degrees
of freedom, is described by local covariant transition rates e.g.
for $q+\bar{q}$ fusion to a meson $m$ of four-momentum $p=
(\omega, {\bf p})$ at space time point $x=(t,{\bf x})$:
\begin{eqnarray}
&&\phantom{a}\hspace*{-5mm} \frac{d N_m(x,p)}{d^4x d^4p}= Tr_q Tr_{\bar q} \
  \delta^4(p-p_q-p_{\bar q}) \
  \delta^4\left(\frac{x_q+x_{\bar q}}{2}-x\right) \nonumber\\
&& \times \omega_q \ \rho_{q}(p_q)
   \  \omega_{\bar q} \ \rho_{{\bar q}}(p_{\bar q})
   \ |v_{q\bar{q}}|^2 \ W_m(x_q-x_{\bar q},p_q-p_{\bar q})
 \times N_q(x_q, p_q) \
  N_{\bar q}(x_{\bar q},p_{\bar q}) \ \delta({\rm flavor},\, {\rm color}).
\label{trans}
\end{eqnarray}
In (\ref{trans}) we have introduced the shorthand notation $Tr_j =
\sum_j \int d^4x_j d^4p_j/(2\pi)^4$ where $\sum_j$ denotes a
summation over discrete quantum numbers (spin, flavor, color);
$N_j(x,p)$ is the phase-space density of parton $j$ at space-time
position $x$ and four-momentum $p$.  In Eq. (\ref{trans})
$\delta({\rm flavor},\, {\rm color})$ stands symbolically for the
conservation of flavor quantum numbers as well as color neutrality
of the formed hadron $m$ which can be viewed as a color-dipole or
'pre-hadron'.  Furthermore, $v_{q{\bar q}}(\rho_p)$ is the
effective quark-antiquark interaction  from the DQPM  defined by
Eq. (31)  of Ref. \cite{Cassing07} as a
function of the local parton ($q + \bar{q} +g$) density $\rho_p$
(or energy density). Furthermore, $W_m(x,p)$ is the phase-space
distribution of the formed 'pre-hadron'. It is taken as a Gaussian
in coordinate and momentum space with width $\sqrt{<r^2>}$ = 0.66
fm (in the rest frame) which corresponds to an average rms radius
of mesons.  The width in momentum space is fixed by the
uncertainty principle, i.e. $\Delta x \Delta p = 1$ (in natural
units).  Related transition rates (to Eq. (\ref{trans})) are
defined for the fusion of three off-shell quarks ($q_1+q_2+q_3
\leftrightarrow B$) to  color neutral baryonic ($B$ or $\bar{B}$)
resonances of finite width (or strings) fulfilling energy and
momentum conservation as well as flavor current conservation
\cite{CasBrat}.

\section{Hadronization of an expanding partonic fireball}

We now turn to actual results from PHSD for the model case of an
expanding partonic fireball at initial temperature $T=1.7\  T_c$
($T_c$= 0.185 GeV) with quasiparticle properties and four-momentum
distributions determined by the DQPM at temperature $T$ = 1.7
$T_c$. The initial distribution for quarks, antiquarks and gluons
in coordinate space is taken as a Gaussian ellipsoid with a
spatial eccentricity
\begin{equation}
\epsilon =\langle y^2-x^2\rangle/\langle y^2 + x^2\rangle
\end{equation}
and $<z^2> = <y^2>$ in order to allow for the buildup of elliptic
flow (as in semi-central nucleus-nucleus collisions at
relativistic energies). In order to match the initial
off-equilibrium strange quark content in relativistic $pp$
collisions the number of $s$ (and $\bar{s}$ quarks) is assumed to
be suppressed by a factor of 3 relative to the abundance of $u$
and $d$ quarks and antiquarks. In this way we will be able to
investigate additionally the question of strangeness
equilibration. For more details of the numerical treatment we
refer the reader to Ref. \cite{CasBrat}.

\begin{figure}[t]
\centering
\includegraphics*[width=150mm]{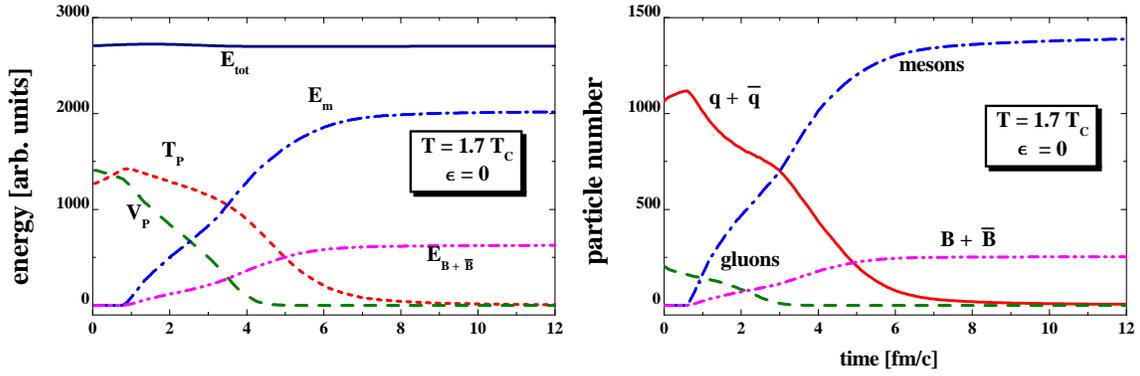}
\caption{(Color online)
L.h.s.: Time evolution of the total energy $E_{tot}$ (upper solid
line), the partonic contributions from the interaction energy
$V_p$ and the energy of time-like partons $T_p$ in comparison to
the energy contribution from formed mesons $E_m$ and baryons (+
antibaryons) $E_{B+{\bar B}}$.  R.h.s:  Time evolution in the
parton, meson and baryon number for an expanding partonic fireball
at initial temperature $T=1.7\  T_c$ with initial eccentricity
$\epsilon = 0$.} \label{fig1}
\end{figure}

In Fig. \ref{fig1} (l.h.s.) we show the energy balance for the
expanding system at initial temperature $T= 1.7 T_c$ and
eccentricity $\epsilon$ =0. The total energy $E_{tot}$ (upper
line) - which at $t=0$ is given by (\ref{ener}) integrated over
space - is conserved within 3\% throughout the partonic expansion
and hadronization phase such that for $t > 8$ fm/c it is given
essentially by the energy contribution from mesons and baryons
(+antibaryons). The initial energy splits into the partonic
interaction energy $V_p$ (cf. Eq. (19) in Ref. \cite{Cassing07})
and the energy of the time-like (propagating) partons
\begin{equation} \label{TK}
T_p = \sum_i \sqrt{p^2_i + M^2_i(\rho_p)}
\end{equation}
with fractions determined by the DQPM \cite{Cassing07}. In Eq.
(\ref{TK}) the summation over $i$ runs over all testparticles in an
individual run. The hadronization mainly proceeds during the time
interval 1 fm/c $< t < $ 7 fm/c (cf.  r.h.s of Fig. \ref{fig1}
where the time evolution of the $q, \bar{q},g$, meson and baryon
(+antibaryon) number is displayed). As one observes from Fig. \ref{fig1}
on average the number of
hadrons from the resonance or 'string' decays is larger than the
initial number of fusing partons.

\begin{figure}[t]
\centering \includegraphics*[width=85mm]{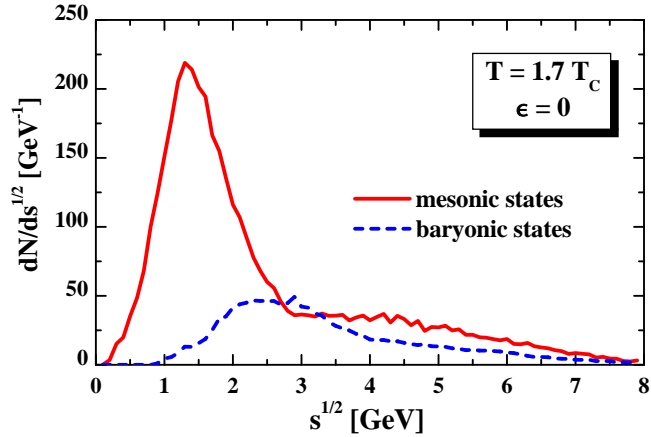}
\caption{(Color online) The invariant mass distribution for fusing $q
\bar{q}$ pairs (solid line) as well as $qqq$ (and
$\bar{q}\bar{q}\bar{q}$) triples (dashed line) that lead to the
formation of final hadronic states for an expanding partonic fireball
at initial temperature $T=1.7\  T_c$ with initial eccentricity
$\epsilon = 0$.} \label{fig2a}
\end{figure}

 In order to shed some light on
the hadronization process in PHSD we display in Fig. \ref{fig2a}
the invariant mass distribution of $q \bar{q}$ pairs (solid line)
as well as $qqq$ (and $\bar{q}\bar{q}\bar{q}$) triples (dashed
line) that lead to the formation of final hadronic states. In
fact, the distribution for the formation of baryon (antibaryon)
states starts above the nucleon mass and extends to  high
invariant mass covering the nucleon resonance mass region as well
as the high mass continuum (which is treated by the decay of
strings within the JETSET model \cite{JETSET}). On the
'pre-mesonic side the invariant-mass distribution starts roughly
above the two-pion mass and extends up to continuum states of high
invariant mass (described again in terms of string excitations).
The low mass sector is dominated by $\rho$, $a_1$, $\omega$ or
$K^*, \bar{K}^*$ transitions etc. The excited 'pre-hadronic'
states  decay to two or more 'pseudoscalar octet' mesons such that
the number of final hadrons is larger than the initial number of
fusing partons.
Accordingly,  the hadronization process in PHSD leads to an increase of
the total entropy and not to a decrease as in case of coalescence
models \cite{Koal1,Koal2}. This is a direct consequence of the finite
(and rather large) dynamical quark and antiquark masses as well as
mean-field potentials which - by energy conservation - lead to
'pre-hadron' masses well above those for the pseudo-scalar meson octett
or the baryon octett, respectively. This solves the entropy problem in
hadronization in a natural way and is in accordance with the second law
of thermodynamics!

The parton dynamics itself is governed by their propagation in the
time-dependent mean-field $U_p(\rho_p)$\footnote{Analoguous
mean-field phenomena are well known from studies on $K^\pm$,
$\rho$-meson and nucleon flow at SIS energies \cite{a1,a2,a3}.}
which is adopted in the parametrized form (as a function of the
parton density $\rho_p$) given by Eq. (29) in Ref.
\cite{Cassing07}. Since the mean-field $U_p$ is repulsive the
partons are accelerated during the expansion phase on expense of
the potential energy density $V_p$ which is given by the integral
of $U_p$ over $\rho_p$ (cf. Sec. 3 in Ref. \cite{Cassing07}).
The interaction rates of the partons are determined by effective
cross sections which for  $gg$ scattering have been determined in
Ref. \cite{Andre} as a function of $T/T_c$. The latter are
re-parametrized in the actual calculation as a function of the
parton density using the available dependence of $\rho_p(T)$ on
the temperature $T$ from the DQPM.
 The channels $q \bar{q} \rightarrow g
$ are described by a relativistic Breit-Wigner cross section which
is determined by the actual masses of the fermions, the invariant
energy $\sqrt{s}$ and the resonance parameters of the gluon (from
the DQPM). In this case a further constraint on flavor neutrality
and open color is employed. The gluon decay to a $u \bar{u}, d
\bar{d}$ or $s\bar{s}$ pair is fixed by detailed balance. Further
channels are $gg \leftrightarrow g$ which are given by
Breit-Wigner cross sections (with the gluon resonance parameters)
and detailed balance, respectively.

It is also interesting to have a look at the final hadronic particle ratios
$K^+/\pi^+$, $p/\pi^+$, $\Lambda/K^+$ etc. (after hadronic decays)
which are shown in Table 1. The latter ratios are compared to the
grand-canonical statistical hadronization model (SM)
\cite{PBM,PBM2} at baryon chemical potential $\mu_B = 0$. For
$\mu_B = 0$ the particle ratios depend only on temperature $T$ and
one may fix a freeze-out temperature, e.g., by the proton to
$\pi^+$ ratio. A respective comparison is given also in Table 1
for $T$ = 160 MeV and 170 MeV for the SM which demonstrates that
the results from PHSD are close to those from the SM for $T
\approx$ 170 MeV. This also holds roughly for the $\Lambda/K^+$
ratio. On the other hand the $K^+/\pi^+$ ratio only smoothly
depends on the temperature $T$ and measures the amount of
strangeness equilibration. Recall that we initialized the system
with a relative strangeness suppression factor of 1/3. The
deviation from the SM ratio by about 13\% indicates that
strangeness equilibration is not fully achieved in the
calculations. This is expected since the partons in the surface of
the fireball hadronize before chemical equilibration may occur.

\begin{table}[h]
\begin{center}
\begin{tabular}{|c|c|c|c|} \hline
\hspace*{3.0cm} &~~~~$p/\pi^+$~~~  & ~~~$\Lambda/K^+$~~~   & ~~~$K^+/\pi^+$~~~~   \\
 \hline
PHSD  &    0.086    & 0.28            &   0.157         \\ \hline
SM $T =160$~MeV &   0.073   & 0.22   &  0.179 \\          \hline
SM $T =170$~MeV &   0.086  & 0.26   &  0.180 \\
 \hline
  \end{tabular}
\end{center}
\caption{Comparison of particle ratios from PHSD with the
statistical model (SM)  for $T$= 160 MeV and 170 MeV.}
\end{table}

The agreement between the PHSD and SM results for the baryon to
meson ratio in the strangeness S=0 and S=1 sector may be explained
as follows:  Since the final hadron formation dominantly proceeds
via resonance and string formation and decay - which is
approximately a micro-canonical statistical process  - the average
over many hadronization events with different energy/mass and
particle number (in the initial and final state) leads to a
grand-canonical ensemble. The latter (for $\mu_B = 0$) is only
characterized by the average energy or an associated Lagrange
parameter $\beta=1/T$.

\begin{figure}[t]
 \includegraphics*[width=85mm]{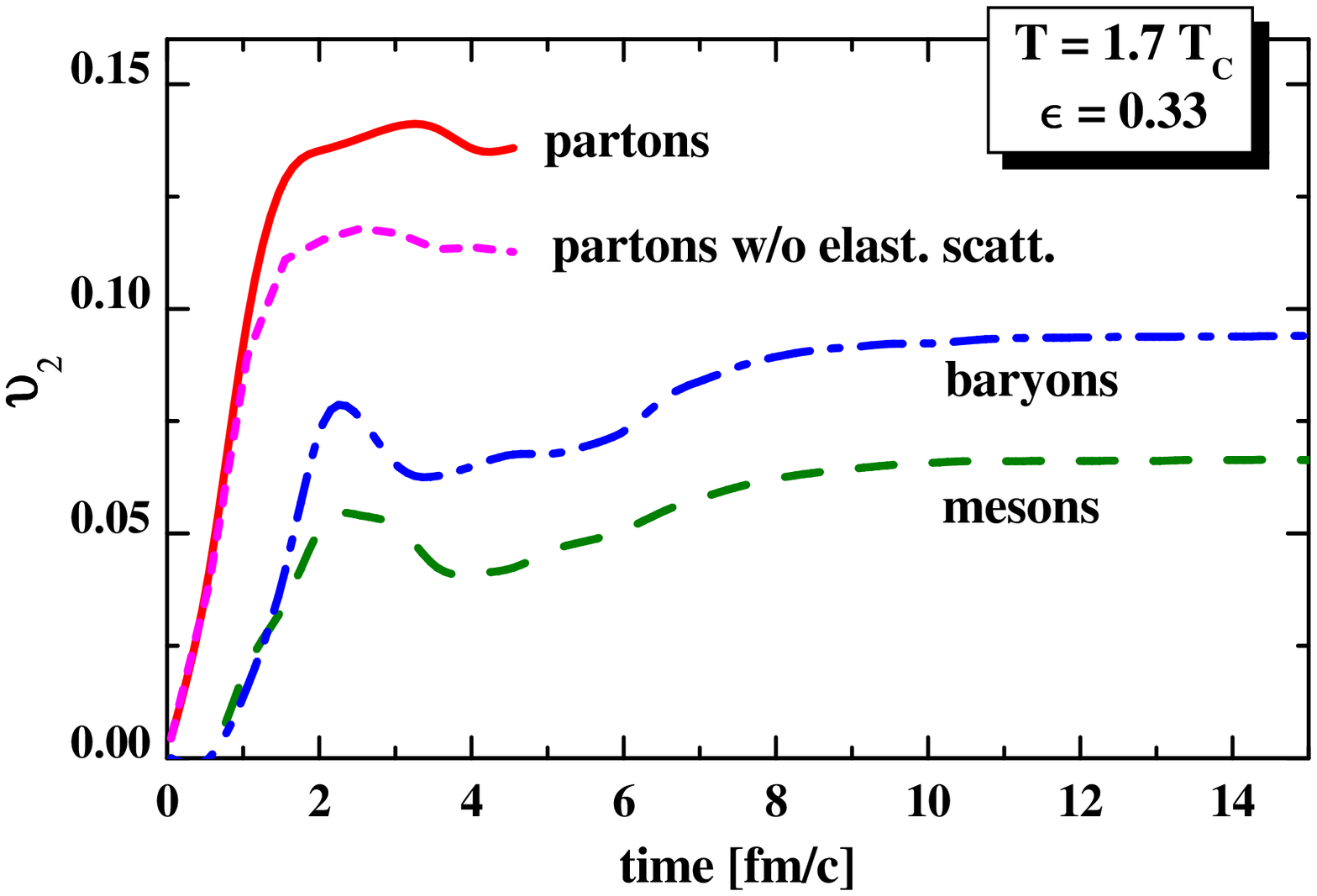} \includegraphics*[width=85mm]{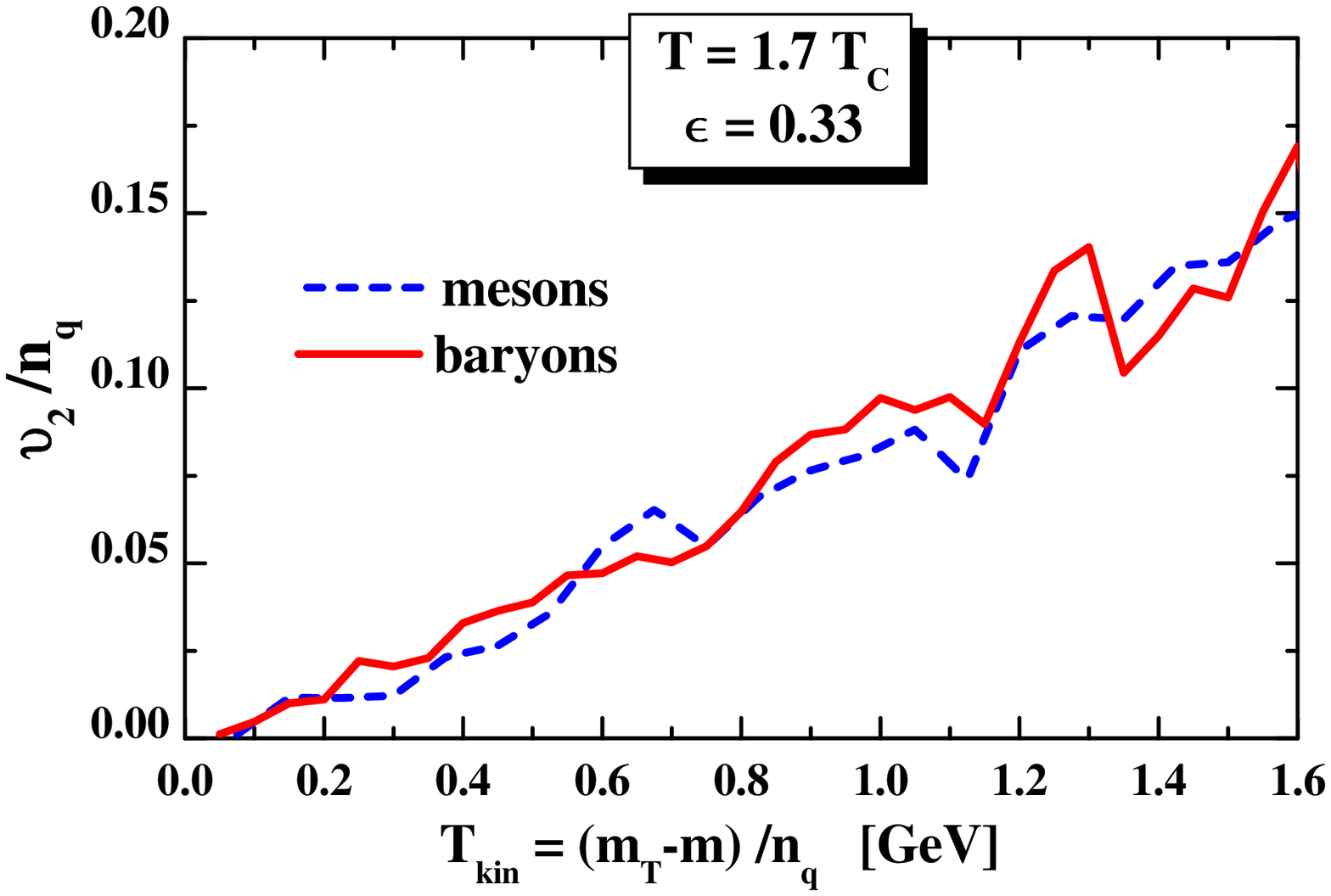} \caption{(Color
online) L.h.s.: Time evolution of the elliptic flow $v_2$ for
partons and hadrons for the initial spatial eccentricity $\epsilon
=0.33$ for an expanding partonic fireball at initial temperature
$T=1.7\ T_c$. R.h.s.: The elliptic flow $v_2$ - scaled by the number
of constituent quarks $n_q$ - versus the transverse kinetic energy
divided by $n_q$ for mesons (dashed line) and baryons (solid
line).}
 \label{fig2}
\end{figure}

Of additional interest are the collective properties of the
partonic system during the early time evolution. In order to
demonstrate the buildup of elliptic flow we show in Fig.
\ref{fig2} (l.h.s.) the time evolution of \begin{equation} v_2 =
\left\langle (p_x^2 - p_y^2)/(p_x^2 + p_y^2) \right\rangle
\end{equation} for partons (solid line), mesons (long dashed line)
and baryons (dot-dashed line) for an initial eccentricity
$\epsilon = 0.33$. As seen from Fig. \ref{fig2} the partonic flow
develops within 2 fm/c and the hadrons  essentially pick up the
collective flow from the accelerated partons. The hadron $v_2$ is
smaller than the maximal parton $v_2$ since by parton fusion the
average $v_2$ reduces and a fraction of hadrons is formed early at
the surface of the fireball without a strong acceleration before
hadronization.  It is important to point out that in PHSD the
elliptic flow of partons predominantly stems from the gradients of
the repulsive parton mean-fields (from the DQPM) at high parton
(energy) density. To demonstrate this statement we show in Fig.
\ref{fig2} (r.h.s.) the result of a simulation without elastic partonic
rescattering processes by the short dashed line.

We note in passing that the final hadron $v_2$ increases linearly
with the initial eccentricity $\epsilon$ and indicates that the
ratio $v_2/\epsilon$ is practically constant ($\approx
0.2$)\cite{CasBrat} as in ideal hydrodynamics. Accordingly the
parton dynamics in PHSD are close to ideal hydrodynamics. This
result is expected since the ratio of the shear viscosity $\eta$
to the entropy density $s$ in the DQPM is on the level of $\eta/s
\approx $ 0.2 \cite{Andre} and thus rather close to the lower
bound of $\eta/s = 1/(4 \pi)$.

A further test of the PHSD hadronization approach is provided by the
 'constituent quark number scaling' of the elliptic flow $v_2$ which
 has been observed experimentally
in central Au+Au collisions at RHIC \cite{STARS,Star08}. In this
respect we plot $v_2/n_q$ versus the transverse kinetic energy per
constituent parton,
\begin{equation} T_{kin} = \frac{m_T-m}{n_q} \ , \end{equation}
with $m_T$ and $m$  denoting the transverse mass and actual mass,
respectively. For mesons we have $n_q = 2$ and for
baryons/antibaryons $n_q=3$. The results for the scaled elliptic
flow are shown in Fig. \ref{fig2} (r.h.s.) for mesons
and baryons  and suggest an approximate scaling. We
note that the scaled hadron elliptic flow $v_2/n_q$ does not
reflect the parton $v_2$ at hadronization and is significantly
smaller. Due to the limited statistics especially in the baryonic
sector with increasing $p_T$ this issue will have to be
re-addressed with high statistics in the actual heavy-ion case
where the very early parton $p_T$ distribution also shows
'power-law' tails.

\section{Summary}
In summary, the expansion dynamics of an anisotropic partonic
fireball has been studied within the PHSD approach which includes
dynamical local transition rates from partons to hadrons
(\ref{trans}) (and vice versa). It shows collective features as
expected from ideal hydrodynamics in case of strongly interacting
systems. The hadronization process conserves four-momentum and all
flavor currents and slightly increases the total entropy  since
the 'fusion' of rather massive partons dominantly leads to the
formation of color neutral strings or resonances that decay
microcanonically to lower mass hadrons. This solves the entropy
problem associated with the simple coalescence model!

We find that the hadron abundancies and baryon to meson ratios are
compatible with those from the statistical hadronization model
\cite{PBM,PBM2} - which describes well particle ratios from AGS to
RHIC energies - at a freeze-out temperature of about 170 MeV. Our
calculations show that the hadron elliptic flow is essentially
produced in the early partonic stage where also the strong
repulsive parton mean-fields contribute to a large extent.

\phantom{a}\vspace{0mm} The authors  like to thank A. Andronic for
providing the SM results in Table 1.  Furthermore, they are
grateful to O. Linnyk for valuable discussions.

%------------------------------------------------------------------------

\end{document}